# Edge conductivity in PtSe$_2$ nanostructures

Roman Kempt, Agnieszka Kuc, Thomas Brumme, Thomas Heine*

Roman Kempt, Dr. Thomas Brumme, Prof. Dr. Thomas Heine
Chair of Theoretical Chemistry, Technische Universität Dresden
Bergstrasse 66, 01069 Dresden, Germany
E-mail: thomas.heine@tu-dresden.de

Dr. Agnieszka Kuc, Prof. Dr. Thomas Heine
Helmholtz-Zentrum Dresden-Rossendorf, Institute of Resource Ecology
Permoserstrasse 15, 04318 Leipzig, Germany

Prof. Dr. Thomas Heine
Department of Chemistry
Yonsei University, Seodaemun-gu, Seoul 120-749, Republic of Korea

## Abstract

PtSe$_2$ is a promising 2D material for nanoelectromechanical sensing and photodetection in the infrared regime. One of its most compelling features is the facile synthesis at temperatures below 500 °C, which is compatible with current back-end-of-line semiconductor processing. However, this process generates polycrystalline thin films with nanoflake-like domains of 5 to 100 nm size. To investigate the lateral quantum confinement effect in this size regime, we train a deep neural network to obtain an interatomic potential at DFT accuracy and use that to model ribbons, surfaces, nanoflakes, and nanoplatelets of PtSe$_2$ with lateral widths between 5 to 15 nm. We determine which edge terminations are the most stable and find evidence that the electrical conductivity is localized on the edges for lateral sizes below 10 nm. This suggests that the transport channels in thin films of PtSe$_2$ might be dominated by networks of edges, instead of transport through the layers themselves.

## Introduction
PtSe$_2$ is a candidate material for the next generation of electromechanical and optical sensors in nanoscale devices.[1,2] It features a large negative Gauge Factor (GF) in electromechanical pressure sensors of up to -84,[3–5] low contact resistivities and large carrier mobilities between 625 and 1500 cm$^2$ V$^{-1}$ s$^{-1}$,[4,6–9] and long-term stability of up to years in presence of air and moisture.[4,10–13] The low thermal budget of 450-500 °C during synthesis, for example via thermally assisted conversion (TAC),[5,7,14] allows PtSe$_2$ films to be directly grown on centimeter-scale silicon wafers and even plastics with different topographies.[3,7] This is ideal to directly integrate PtSe$_2$ into devices, such as piezoresistive sensors,[1,15,16] photonic circuits,[17–19] and memristors.[20]
There are still challenges in consistently growing high-quality PtSe$_2$ films, such as controlling homogeneity,[6,7] layer orientation,[8] continuity,[6,21] and layer thickness.[5,7,22,23] Often, the resulting films are polycrystalline with fused, nanoflake-like domains between 10 to 50 nm and thicknesses between 6 to 8.5 nm.[5–7,12] Recently, chemical vapor deposition (CVD) with a metal-organic precursor was shown to yield domains of up to 300 nm in size.[23]

Controlling these synthesis factors is crucial to obtain $PtSe_2$ devices with reproducible performance because they strongly affect the electronic properties of the $PtSe_2$ film. For example, varying the layer thickness from two to three layers gives rise to a semiconductor-to-semimetal transition,[24–26] point- and edge-defects cause magnetic behavior,[27–29] and stacking faults lead to semiconducting instead of semimetallic behavior.[30,31] All of those factors may contribute to varying observations of $PtSe_2$ characteristics, such as mobilities below 1 $cm^2$ $V^{-1}$ $s^{-1}$,[32] p- or n-type behavior depending on the selenization process,[33] and a 35% reduction of the cross-plane thermal conductivity due to polycrystallinity.[34]

In this study, we investigate how the electrical conductivity depends on the lateral width of $PtSe_2$ nanostructures, namely ribbons, surfaces, nanoflakes and nanoplatelets, with different edge terminations using density-functional theory (DFT). The models required to study such systems contain up to a few thousand heavy atoms. To this end, we train a deep neural network to obtain an interatomic potential for $PtSe_2$ nanostructures. The training data includes three different stacking phases in their bulk form and for up to nine layers (from our previous work)[30] and differently terminated ribbon and surface models (see **Figure 1a-d**). We employ the DeePMD-kit framework[35] which implements the smooth version of the Deep Potential (DP) model by E et al.[36,37] The DP model has been successfully used to represent the Potential Energy Surface (PES) of complex systems in materials science, such as the formation of defects from radiation exposure under fusion reactor conditions[38] or the behavior of water confined between graphene nanocapillaries.[39] Furthermore, it has been used to explain the large piezo- and pyroelectric coefficients of zirconia at high temperatures,[40] the large lattice thermal conductivity of $Bi_2Te_3$ with strong anharmonicity,[41] and even moiré phonons in graphene.[42] The DP model has been applied to systems containing more than 100 million atoms with effectively the same accuracy as the underlying training method.[43]

Using the DP model, we find that the stability of different edge terminations depends strongly on temperature and is influenced by the number of layers of $PtSe_2$. We calculate the band structures and Boltzmann conductivities of ribbons, surfaces, nanoflakes and nanoplatelets of $PtSe_2$ for lateral widths between 5 and 15 nm using DFT, with the largest structure containing more than 1200 heavy atoms in periodic boundary conditions along the stacking direction. All structures have been made available at the NOMAD repository as a data set.[44]

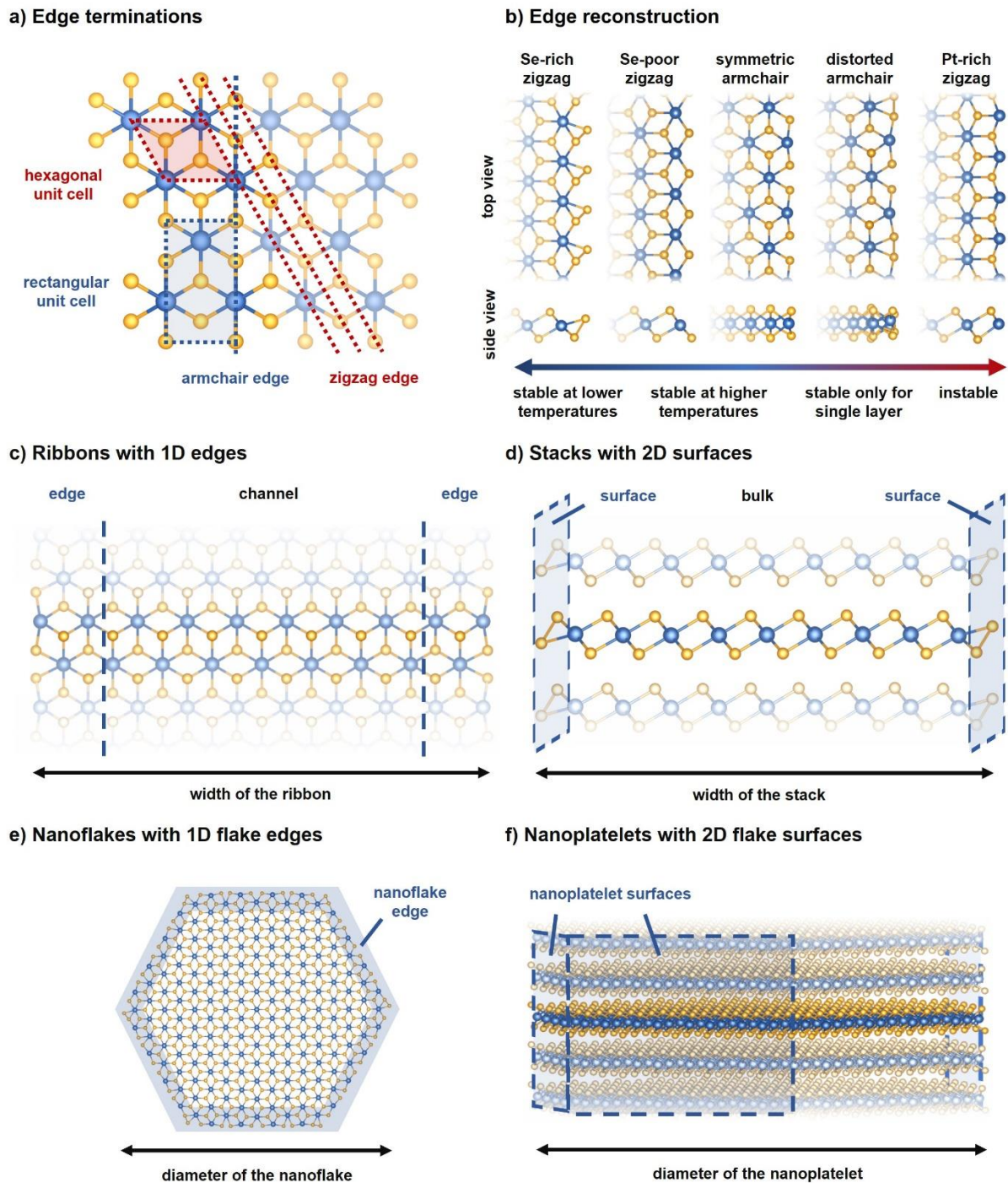

**Figure 1** – **a)** Different unit cell choices to model PtSe$_2$ and the cutting vectors to obtain the armchair and zigzag edges with different terminations. **b)** Edge reconstruction of the five edge terminations after structural relaxation and their order of stability. **c)** Ribbon model with a single periodic axis to investigate one-dimensional edges. **d)** Stack model with two periodic axes to investigate two-dimensional surfaces with different edge terminations. **e)** Hexagonal nanoflake model without periodicity to investigate edge states. **f)** Hexagonal nanoplatelet model with one periodic axis to investigate two-dimensional surface states on all six sides.

## Results and Discussion

### Training Data Set and Settings

We employ a data set that includes PtSe$_2$ structures in the octahedrally coordinated 1T$^O$, 2H$^O$, and 3R$^O$ stacking phases as discussed in our previous work,[30] as well as bulk phases, ribbons, and surfaces.[44] All trajectories have been calculated at the same level of theory (see **Methods** for details). The DP model was fed with energies, forces and stresses (where available) from relaxation trajectories, phonon calculations, elastic deformations, and Molecular Dynamics (MD) simulations. The composition of the data set is summarized in **Table S1**. The largest fraction of frames stems from MD (NVT ensemble) simulations (77 %), which were performed on different unit cell sizes up to 700 K. The 1T$^O$-phase is the largest fraction of the training data structure type (69 %) because it is the one that defines edges as shown in **Figure 1**. For reasons discussed below, we do not include other polytypes, such as the 3T$^O$, 3A$^O$ and 6R$^O$ phases.[30] At the current stage, the model includes only cell stresses from bulk unit cells, as we fix the unit cell parameters of edges and surfaces at their corresponding few-layer or bulk counterparts. An exemplary training parameter input is shown in **Figure S1**. The data set is rather small with approximately 12700 frames, which we use entirely for training. We validate the performance of the DP model not just on the as-obtained energies and forces, but on derived quantities, such as bond lengths and phonon spectra.

### Accuracy of the DP model

To verify the validity of the DP model, we analyze how the energy and force root mean square error (RMSE) distributes over the training data set and how accurately it predicts bond lengths and phonon spectra (see **Figure 2 a-c**). We find that the DP model represents the potential energy surface of the system well, yielding a weighted RMSE of 3.5 meV atom$^{-1}$ for the energies and a weighted RMSE of 18.9 meV atom$^{-1}$ Å$^{-1}$ for the forces (see **Figure 2 a**). The prediction of atomic forces is better than the prediction of energies because there is more information to train from. The forces are atom-resolved (three numbers per atom), whereas the total energy is a single value averaged over the whole system. We find that the energy prediction error is larger (between 5 to 10 meV atom$^{-1}$) for systems with single unit cells (i.e., primitive units) than for systems in supercell representation, likely because the training data includes more supercell calculations than calculations in the primitive unit cell. An additional effect can be observed in single-layered PtSe$_2$, where the predicted total energy decreases with increasing supercell size (see **Figure S2**) because the structure is allowed to undergo distortions, such as wrinkling, which are not possible in a single unit cell. Thus, the DP model underestimates the total energy of single-layer PtSe$_2$ in its primitive unit by 19 meV atom$^{-1}$. We conclude that the DP model should only be used for sufficiently large unit cells (above 50 atoms), which is the case in this study. The error of the force prediction becomes largest for structures which deviate from the ideal 1:2 Pt:Se ratio, which are Pt-rich or Se-rich edge models. This error might be further reduced by adding additional training data.

The resulting bond lengths are reproduced within 2% accuracy of the DFT value (see **Figure 2 b**). We observe that larger errors occur for the prediction of Se-Se dimer bond lengths, which form only in the Se-rich zigzag edge and the single-layer armchair edge. This might be remedied by adding more training data for those specific edges and for the pure bulk phases. There are larger difficulties in the prediction of the interlayer distance by the DP model, which tends to underestimate interlayer distances between 0.5 and 1.4 % (see **Figure 2 b**) compared to the DFT value. This error cannot be easily remedied by adding more training data but stems from how the

## a) Energy and force prediction
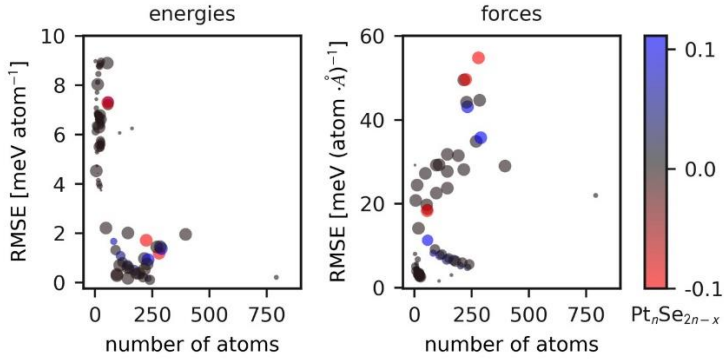

## b) Bond length prediction
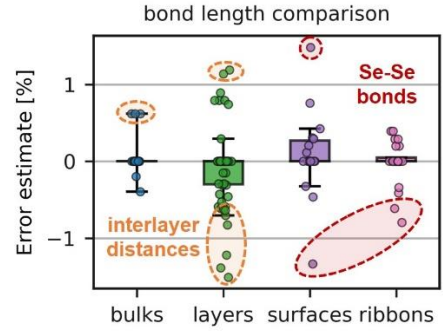

## c) Comparison of bulk phonon spectra
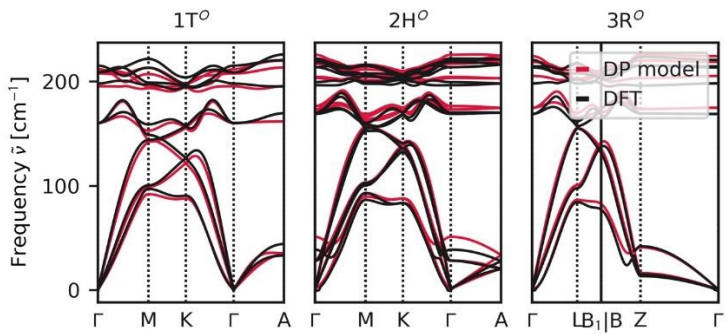

## d) Nanoflake stability
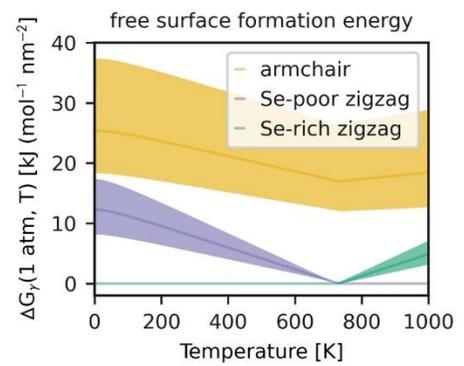

**Figure 2** – **a)** Energy and force prediction of the DP model as obtained by the *dp test* utility over the training data set. Each circle corresponds to a subset of input frames with the same chemical formula. The size of the circle corresponds to the size of the subset and the color indicates the deviation from the ideal stoichiometry of 1:2 in PtSe$_2$. **b)** Relative bond length error compared between structures after relaxation by the DP model and from DFT. Each dot represents the comparison between the smallest, largest or mean bond length of the same type of bond. **c)** Bulk phonon spectra of three stacking phases obtained from DFT and the DP model. **d)** Relative free surface formation energy of the differently terminated nanoflake models. The width of the bands corresponds to energy differences obtained for different lateral diameters of the nanoflakes.

DP model is trained. The DP model predicts energies of single atoms embedded in a local bonding environment, represented by *environment matrices* $(\mathcal{R}_i)_j$:[35,38]

$$(\mathcal{R}_i)_j = s(|\boldsymbol{r}_{ij}|) \times \left(\frac{x_{ij}}{|\boldsymbol{r}_{ij}|}, \frac{y_{ij}}{|\boldsymbol{r}_{ij}|}, \frac{z_{ij}}{|\boldsymbol{r}_{ij}|}\right) \text{ with } s(|\boldsymbol{r}_{ij}|) = 0 \text{ for } |\boldsymbol{r}_{ij}| > r_c$$

where $|\boldsymbol{r}_{ij}| = |\boldsymbol{r}_j - \boldsymbol{r}_i|$ and $x_{ij}$, $y_{ij}$, and $z_{ij}$ are the Cartesian components of the relative distance vector to other atoms within a cutoff radius $r_c$. The factor $s(|\boldsymbol{r}_{ij}|)$ is a switching function that smoothly varies from 1 to 0 at this cutoff distance.[38] These environment matrices define the feature matrix of the DNN. Consequently, the DP model can only learn from features which are representable within this cutoff radius. For a layered material, such as PtSe$_2$, a spherical cutoff is not ideal, because one would have to choose a very large radius to include next-layer and next-next-layer interactions. However, training quickly becomes unfeasible for too large cutoff radii, and the prediction may become worse because the feature matrix becomes diluted. Furthermore, there are long-range interactions (such as dielectric screening), which strongly affect the interlayer distance, but cannot be captured by a simple cutoff radius at all. To solve this issue, further developments in DeePMD-kit are required, e.g., anisotropic cutoff radii.

Here, we chose a cutoff radius of 8 Å for training, which means that each individual layer at most sees the next and second next layer (for comparison, the bulk interlayer distance in the $1T^O$ phase at the DFT level is 4.97 Å). This yielded a good compromise to predict accurate interlayer distances and edge configurations. Consequently, we do not include more stacking orders with larger interlayer distances and long-range stacking effects in our training data set (such as $3T^O$, $3A^O$ and $6R^O$).[30] These would require further training data and improved cutoff definitions.

### Edge Stability

The DP model reproduces the phonon spectra of the three stacking orders $1T^O$, $2H^O$ and $3R^O$ in the bulk phase very well, as shown in **Figure 2 c**. All three stacking orders are found to be stable local minima, agreeing with our previous work.[30] Interestingly, the DP model shows a degeneracy of phonon modes at the A-point in the $1T^O$ and $2H^O$ stacking configuration, whereas the DFT calculations do not. This hints at a lack of long-range interaction in the DP model, which might be explained by the cutoff radius.

Using the DP model, we can estimate thermodynamic quantities for very large systems, such as nanoflakes. We argue that nanoflake models are the most realistic approach to determine the stability of edge terminations because they allow for the comparison of systems with similar finite sizes without considering the additional effects of interacting periodic images. Furthermore, they are the best model to consider nanoflake growth, which depends, e.g., on charge localization on the corners of such nanoflakes.[45] Still, one has to choose a thermodynamic model to compare the Gibbs free energies of systems with different compositions and number of atoms. In the TAC process, a pre-deposited Pt film is converted to $PtSe_2$ by pure, vaporized Se powder transported by a carrier gas (e.g., $H_2$).[7] Hence, choosing bulk Pt and bulk Se in their most stable forms is the simplest reference for the (unbalanced) reaction:

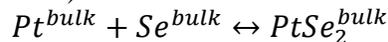

$$Pt^{bulk} + Se^{bulk} \leftrightarrow PtSe_2^{bulk}$$

If we normalize the Gibbs free energy of this reaction over the area of a nanoflake, we obtain a free surface formation energy:[46]

$$G_\gamma(p,T) = \frac{1}{2a}\left(G^{nanoflake}(p,T) - n_{Pt}G^{bulk\,Pt}(p,T) - n_{Se}G^{bulk\,Se}(p,T)\right)$$

Here, $p$ stands for pressure (we chose 1 atm), $T$ for temperature, $a$ for the basal area of the hexagonal flake, $n_{Pt}$ and $n_{Se}$ for the number of Pt and Se atoms in the $PtSe_2$ nanoflake, respectively. The individual terms $G(p,T)$ are the free energies including the internal energy $E_{pot}^{DFT} + U(T)$, the Zero-Point Energy (ZPE) and the entropy $S(T,p)$. The pressure-volume term is included for the molecular nanoflakes, but neglected for the bulk crystals, where the change in volume at this pressure is negligible. The results are shown in **Figure 2 d**, where different values are obtained for nanoflakes with different widths between 5 and 8.7 nm, yielding a range of free formation energies (indicated as bands). Within this model, the Se-rich zigzag termination is determined as the most stable for temperatures below 500 °C, agreeing with the observations of Li et al.[29] For higher temperatures, the Se-poor zigzag termination becomes more favorable. In experiments, Se tends to evaporate from $PtSe_2$ films at higher temperatures, which is not considered in this simple thermodynamic model. As Li et al.[29] pointed out, this likely leads to generally Se-deficient PtSe$_2$ films, which means the Se-poor zigzag edge is more abundant than estimated by this model.

Regarding the Pt-rich zigzag edge (see **Figure 1 b**), we exclude it in further analysis because we observe already in the DFT MD simulation that the edge breaks down under annealing (see **Figure S3**). In experiments, this edge probably requires surface passivation to be stabilized. The Pt-rich edge is the only edge termination where we have observed magnetization, which has been discussed for the Se-poor and Pt-rich edge by Li et al.[29] Similarly, Avsar et al.[27,28] have shown that Pt-vacancies in single-layer PtSe$_2$ give rise to magnetization. We cannot exclude that the other edges also contribute to observed magnetic behavior in PtSe$_2$ films[27,47] because the PBE functional might simply underestimate it. At the current stage, we investigate the electronic properties taking spin-orbit coupling into account, but no other spin polarization.

Generally, we observe that the armchair edge is less stable than the Se-poor and Se-rich zigzag edge (see **Figure 2 d**). The armchair edge undergoes strong surface reconstruction for a single layer of PtSe$_2$, which we refer to as the distorted armchair edge. The distorted armchair edge doubles its periodicity to form alternating Se-Se dimers (see **Figure 1 b, Figure S4**). This does not occur for surfaces, where the symmetric armchair edge is preferred. This might be rationalized by the larger dielectric screening and the larger bulk lattice constant compared to a single layer. This feature is correctly reproduced by the DP model, however, it overestimates the transition barrier between the symmetric and distorted armchair edge for single-layer PtSe$_2$. For example, for a ribbon with 5.52 nm width, the DFT transition barrier is 12.2 kJ mol$^{-1}$ unit$^{-1}$, whereas the DP model predicts a barrier of 18.7 kJ mol$^{-1}$ unit$^{-1}$. The larger barrier predicted by the DP model likely stems from the larger energy prediction error on structures containing Se-Se dimers, which only make up a small fraction of the training data set.

Lastly, we observe wrinkling (or rippling) in single-layered nanoflakes with the Se-poor and Se-rich zigzag edge, but not in nanoflakes with the armchair edge (see **Figure S5**). Such a wrinkling effect has been previously observed for other TMDC monolayers and may lead to a significant reduction of the band gap.[48] We observe that the DP model appears to have a tendency to induce larger wrinkling in such nanostructures than would be obtained from DFT (see **Figure S5**). In nanoribbon models, the effect is small and impacts the electronic properties only to a small amount (see **Figure S6**). In nanoplatelet models, the wrinkling is smaller due to the periodicity along the stacking axis. The wrinkling may contribute to the localization of conductive channels, though, as we show in the next section, we observe such localization for the perfectly flat armchair nanoflake as well.[48]

Lateral Quantum Confinement in Ribbons and Surfaces

Before we discuss the electronic properties of nanoflakes and nanoplatelets, we take a look at the electronic band structures of the corresponding ribbons and surfaces with the same edge termination, focusing on the change of electronic bands when increasing lateral width. In **Figure 3**, we project the electronic band structure onto the atoms belonging to the edge (within a 1 nm region) and the central PtSe$_2$ units which we call channel or bulk (see **Figure 1**). The full set of band structures is shown in **Figures S7-8**. In the case of single-layered ribbons, we observe a clear distinction between edge bands and channel bands. For the armchair and Se-poor zigzag edge, the edge bands reside within the band gap of the PtSe$_2$ channel, as was also observed by Li et al.[29]

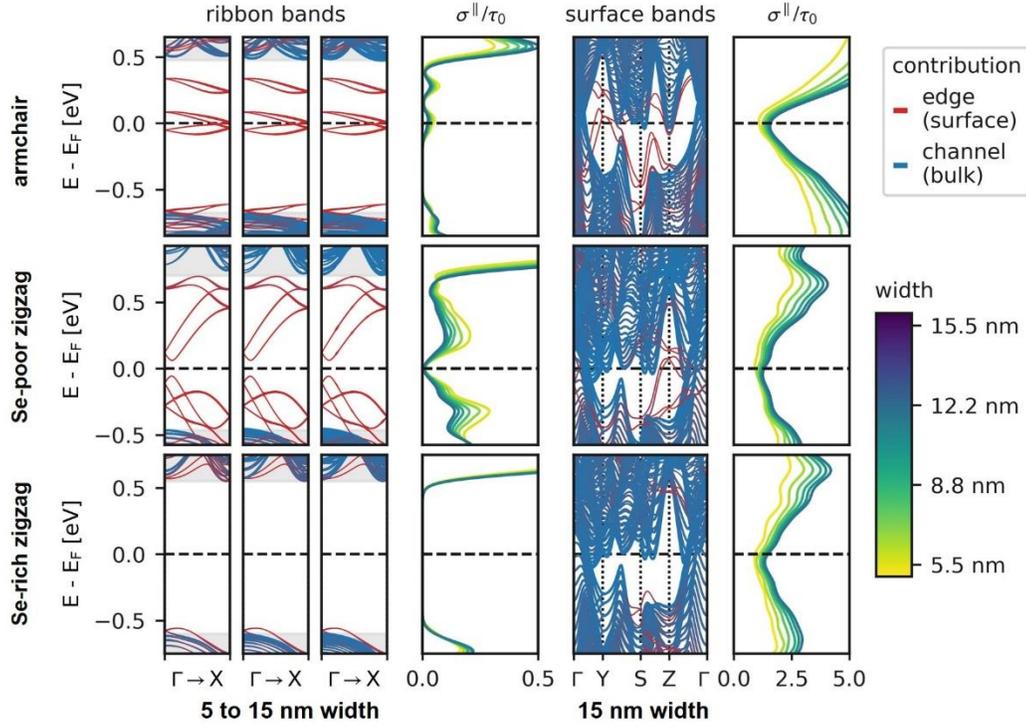

**Figure 3** – Electronic band structures of ribbons and surfaces of PtSe$_2$ projected onto the edge atoms within a region of 1 nm for different lateral widths and different edge terminations. Gray bars indicate the band edges of the PtSe$_2$ channel in ribbons. The Boltzmann conductivity in the constant relaxation time approximation is shown parallel to the edge (or surface) for different lateral widths at 300 K (the unit is $10^{21} \Omega^{-1} m^{-1} s^{-1}$).

The bands of the PtSe$_2$ channel are affected by quantum confinement, whereas the edge bands are not. In the limit of large lateral width, the concentration of edge states vanishes and can be considered dopant states, leaving the band edges of the PtSe$_2$ channel (indicated by gray areas in **Figure 3**). For nanoribbons, we observe that the band gap converges to a value of about 1.2 eV after 10 nm width. The band gap of perfect, single-layered PtSe$_2$ at the PBE level is 1.38 eV. The difference between the two values might be due to wrinkling.[48] The distorted armchair edge is formally metallic, whereas the Se-poor zigzag edge is semiconducting with a small band gap, and the Se-rich zigzag edge is insulating with the edge states residing close to the channel band edges. This does not reflect how the conductivity of the edges behaves. In **Figure 3**, we show the Boltzmann conductivity in the constant relaxation time approximation for different lateral widths.[49] The Boltzmann conductivity is a tensor $\boldsymbol{\sigma}/\tau_0$, where $\tau_0$ is the so-called relaxation time. The relaxation time of a system is in principle unknown and depends on how quickly excited carriers scatter back into the ground state due to, e.g., lattice vibrations. For reference, a typical relaxation time is in the range of $10^{-13}$ to $10^{-15}$ s (e.g., SnSe$_2$)[50] depending on temperature and doping.[49,51] If we assume that the relaxation time for conductivity parallel to the edge does not depend much on the width of the ribbon and the termination of the edge, we can make qualitative comparisons between them. To this end, we project the conductivity tensor onto the lattice vectors (here $b$ and $c$) running parallel to the edges and surfaces:

$$\sigma^\parallel = b^T \boldsymbol{\sigma} b \quad \text{(edges)}$$

$$\sigma^{\parallel} = \frac{1}{2}(b^T \boldsymbol{\sigma} b + c^T \boldsymbol{\sigma} c) \quad \text{(surfaces)}$$

With this, we can show that the distorted armchair edge features very small conductivity, even though it is formally metallic. This can be understood by taking into account that the bands crossing the Fermi level are flat and stem from lone pairs of the Se-Se dimers forming in the distorted armchair edge. For comparison, the symmetric armchair edge, which is instable for single-layer PtSe$_2$, is semiconducting with a higher conductivity near the Fermi level (see **Figure S3**). Consequently, the transition from the symmetric to the distorted armchair edge in single-layer PtSe$_2$ can be seen as a transition from a conducting state to a nearly insulating state. The Se-rich zigzag edge shows insulating behavior in the conductivity, whereas the Se-poor zigzag edge has the highest conductivity near the Fermi level. This conductivity decreases with increasing width of the nanoribbon, possibly because the edge states become more delocalized due to the larger screening of the PtSe$_2$ channel.

The analysis of edge states in the case of two-dimensional bulk surfaces becomes more involved, as shown on the right in **Figure 3**, because PtSe$_2$ undergoes a transition from semiconductor to metal when going from a single layer to bulk.[24,52] Most bands crossing the Fermi level, especially near the Γ-point, stem from bulk PtSe$_2$ states. Hence, the conductivity depends less on the edge termination and generally increases with increasing lateral width. In **Figure S8**, we show how the band structure changes for increasing lateral width of the stack (see **Figure 1 d**). Until ca. 10 nm, new bands approach and cross the Fermi level, whereas after 10 nm, we consider the band structure converged. Only in the case of the armchair and Se-poor zigzag edge, there are surface bands crossing the Fermi level. This indicates that there is large surface conductivity for these two models, whereas the bulk Se-rich zigzag surface has lower conductivity. In the case of the Se-poor zigzag edge, specifically, there are edge bands crossing at the Z-point, which corresponds to the stacking direction. This indicates that interlayer coupling of the Se-poor zigzag edges facilitates out-of-plane surface transport.

### Edge conductivity in PtSe$_2$ nanoflakes and nanoplatelets

After discussing edge bands in ribbons and nanoplatelets, we can explain how these features are represented in single-layered nanoflakes and nanoplatelets. In **Figure 4 a**, we show the integrated Local Density of States (iLDOS) for different edge terminations integrated over the partially occupied states near the Fermi level. These states are partially occupied due to thermal smearing and serve as conductive channels localized on the nanoflake edges, agreeing with the observations in nanoribbons for the Se-poor zigzag and armchair edges. The Se-rich zigzag edge has only very few partially occupied states stemming from excess Se clusters. Disregarding these, the Se-rich zigzag nanoflakes are insulating with band gaps of 1.02, 0.98, and 0.96 eV for diameters of 5.62, 7.13 and 8.64 nm, respectively. These band gap values are smaller than in pristine monolayer PtSe$_2$, which might be attributed to the wrinkling effect.[48]

Similar to the bulk surfaces, when nanoflakes are stacked together to form vertical nanoplatelets, they undergo a transition from semiconductor to metal (see **Figure 4 b**). The contribution of edge atoms to the band structure is much larger in nanoplatelets with a smaller diameter than in comparable ribbon or surface models, because the number of edge atoms makes up a larger fraction. For large nanoplatelet models, the number of edge atoms is proportional to the radius $r$, whereas the number of atoms in the PtSe$_2$ channel is proportional to the square of the radius, $r^2$. Hence, for the smallest nanoplatelets with 5.2 to 5.62 nm width, we find that the states around the

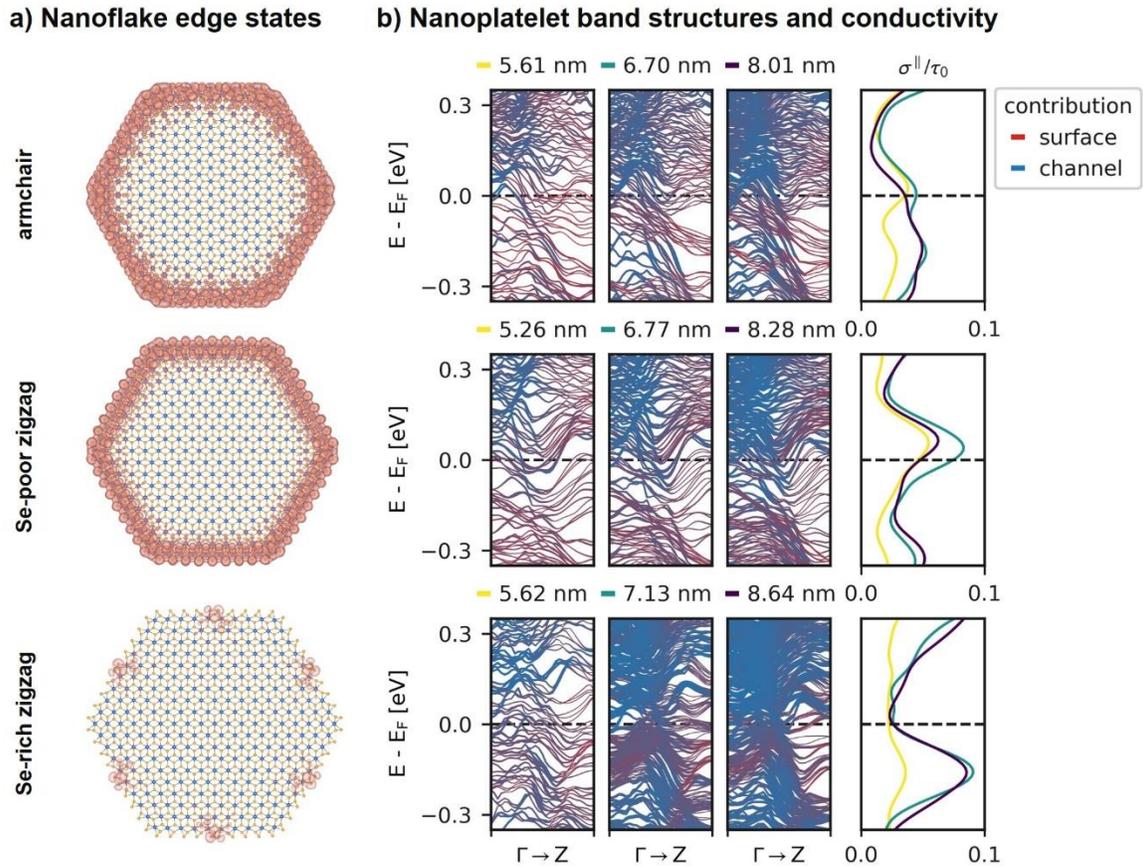

**Figure 4** – a) Integrated Local Density of States of the partially occupied orbitals at an isosurface value of 0.05 for the differently terminated nanoflakes at medium diameter. b) Electronic band structure projected on the surfaces of the hexagonally shaped nanowires for different diameters and their conductivity at 300 K with constant relaxation time approximation (the unit is $10^{21} \Omega^{-1} m^{-1} s^{-1}$).

Fermi level have large contributions from the edge atoms for all three edge terminations. Only for the largest nanoplatelets with 8.01 to 8.64 nm width, the bands from the PtSe$_2$ channel become continuous as in the corresponding direction $\Gamma \rightarrow Z$ in the surface model (see **Figure 3**). The conductivity, which, in this case, corresponds to the interlayer conductivity parallel to the surfaces of the nanoplatelet, is smaller than for the corresponding surface models by an order of magnitude (see **Figure 4 b**). As discussed for the surface models, the Se-poor zigzag edge features the largest conductivity at the Fermi level, because the coupling of the Se-poor zigzag edges appears to facilitate interlayer transport. The Se-rich zigzag edge features the largest conductivity below the Fermi level because the edge states of the Se-rich zigzag edge are deep below the Fermi level, as seen in the surface band structures in **Figure 3**.

To analyze whether electrical conductivity occurs on the edges or in the PtSe$_2$ channel of nanoplatelets, we consider the integrated Local Density of States, as shown in **Figure 4 a**. We integrate the Local Density of States (LDOS) separately for holes and electrons over the same energy range as shown in the band structures in Figure **4 b** and project it onto the basal plane of a single nanoflake in the nanoplatelet. The result for the largest nanoplatelets is shown in **Figure 5** and for the smaller nanoplatelets in **Figures S9-10**.

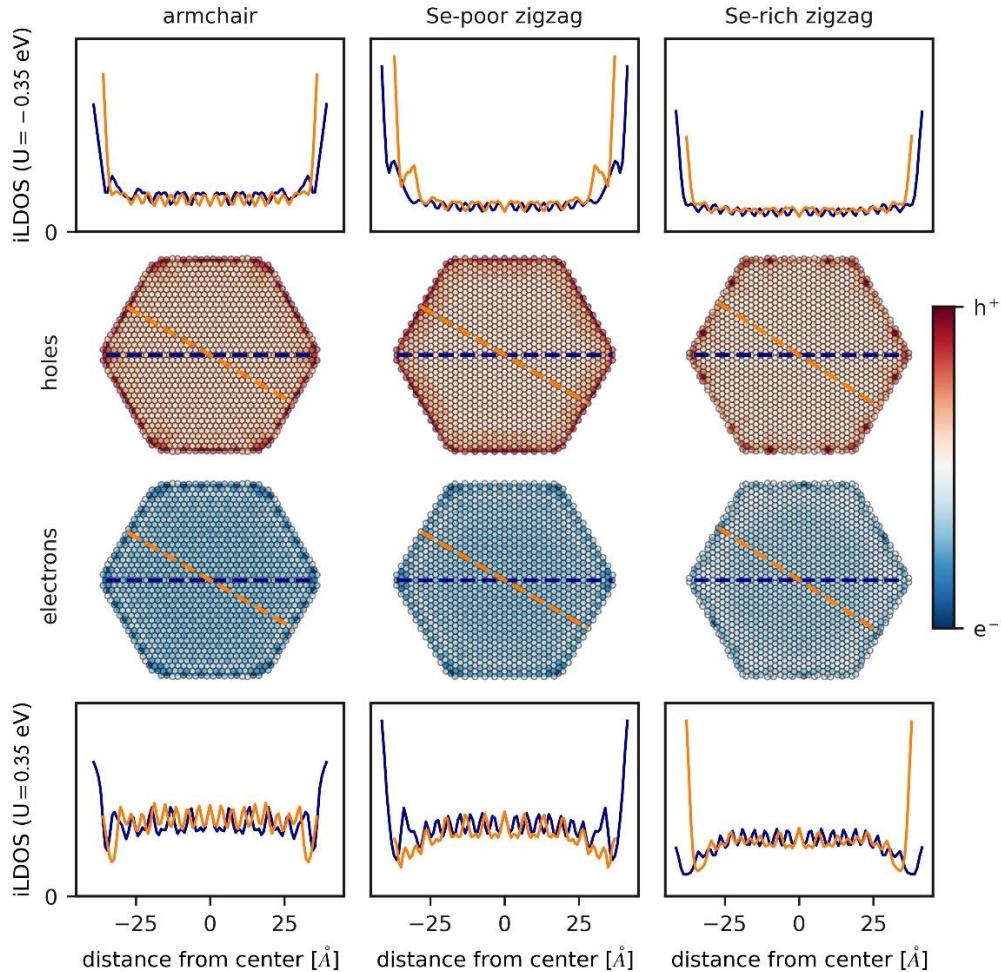

**Figure 5** – Integrated Local Density of States (iLDOS) along line scans and projected onto the basal plane of the nanoplatelets with 8.01 to 8.64 nm lateral width for different edge terminations. The local density of states has been integrated over the holes and electrons in the same energy window as shown in the band structures in **Figure 4**.

For the hole states, we show that the largest iLDOS is localized on the edges for all three edge terminations, which also applies to small, medium-sized, and large nanoplatelets. The iLDOS in the center of the nanoplatelet is smaller, but non-zero, as the nanoplatelets are metallic. This is especially evident in the line scans shown in **Figure 5**, which can be measured experimentally, for example, with scanning-tunneling spectroscopy.[29] For the electron states, on the other hand, the edge states are localized on the corners of the nanoplatelet for the armchair and Se-poor zigzag edge, which agrees with the observation made by Miró et al.[45] for 1T-TiS$_2$ nanoflakes. For the Se-rich zigzag edge, the electrons are rather localized in the center of the nanoplatelet, except for an increase where excess Se clusters have formed. The Boltzmann conductivity is proportional to the available DOS and the band velocities,[49] which are non-zero for the disperse bands shown in **Figure 4**. Hence, we conclude for all three edge terminations in the size regime below 10 nm that the holes are conducted to a significant amount over the surfaces of the nanoplatelets, whereas electrons are conducted along the nanoplatelet corners for the armchair edge and the Se-poor zigzag edge, but not for the Se-rich zigzag edge.

## Conclusions

We investigate $PtSe_2$ nanostructures, including ribbons, surfaces, nanoflakes and nanoplatelets, by means of Density-Functional Theory. To model these systems at sizes comparable to experiments, we train a Deep Neural Network interatomic potential with DeePMD-kit. This allowed us to study nanostructures with different edge terminations, namely the armchair edge and the three different zigzag edges with Se-poor, Se-rich and Pt-rich termination. The interatomic potential accurately reproduces energies, forces and bond lengths of the systems in question with a tendency to underestimate interlayer distances. With the interatomic potential, we determine the stability of the edges in nanoflakes, where we find that the Se-rich zigzag edge is most stable at lower temperatures but competes with the Se-poor zigzag edge at temperatures that are reached during synthesis. We investigate the lateral quantum confinement effect in ribbons, surfaces and nanoplatelets using DFT calculations on the structures obtained from the interatomic potential. The armchair and Se-poor zigzag edges lead to conducting edge states that dominate the electronic structure independent of the lateral width, whereas the Se-rich zigzag edge leads to insulating edge states. This is reflected in the Boltzmann conductivities, where we observe the largest conductivities for the Se-poor zigzag edge in ribbons and nanoplatelets. Particularly, we find that the armchair and Se-poor zigzag edge enhance the interlayer conductivity in form of conductive channels localized on the surfaces and corners of $PtSe_2$ nanoplatelets. We argue that, especially in the size regime below 10 nm, the electrical conductivity of $PtSe_2$ nanoplatelets is edge-dominated, which is crucial for understanding device contacts, catalytic properties, and molecular sensing.

## Data Availability

The training data set has been made available in the NOMAD repository.[44] DeePMD-kit input configuration files and further details can be found in the SI and are available from the author upon reasonable request.

## Methods

For all DFT calculations, including training data and electronic structure calculations, we employ the Fritz Haber Institute Ab Initio Material Simulations (FHI-aims) suite on intermediate and tight tier one basis sets (2020 default).[53] We use *k*-point densities ranging from 12 to 15 points per Å with the PBE functional and added non-local many-body dispersion correction (MBD-nl).[54,55] All electronic structure calculations include spin-orbit coupling (SOC). For phonon calculations and Molecular Dynamics (MDs), we employ phonopy and FHI-vibes.[56,57] MDs are run with the Langevin thermostat at timesteps of 5 fs for 500 steps targeting 700 K as temperature with a friction coefficient of 0.02 as implemented in the Atomic Simulation Environment (ASE).[58] Electrical conductivities are calculated via BoltzTraP2 using an interpolation parameter of 12 and integrated at 300 K without taking doping effects into account (chemical potential equals Fermi level).[49] The DP model was trained using DeePMD-kit with training parameters shown in the Supporting Information.[35–37]

## Supporting Information
Supporting Information is available online or from the author.

## Acknowledgements

This work was financially supported by the German Ministry of Education and Research (BMBF) under the project ForMikro-NobleNEMS (16ES1121). The authors gratefully acknowledge the Gauss Centre for Supercomputing e.V. (www.gauss-centre.eu) for funding this project by providing computing time through the John von Neumann Institute for Computing (NIC) on the GCS Supercomputer JUWELS at Jülich Supercomputing Centre (JSC). The authors are grateful to the Center for Information Services and High Performance Computing at TU Dresden for providing its facilities for the training of the DP model. The authors also thank DFG within CRC 1415 project for support.

# Edge conductivity in PtSe$_2$ nanostructures

## Supporting Information

Roman Kempt, Agnieszka Kuc, Thomas Brumme, Thomas Heine*

Roman Kempt, Dr. Thomas Brumme, Prof. Dr. Thomas Heine
Chair of Theoretical Chemistry, Technische Universität Dresden
Bergstrasse 66, 01069 Dresden, Germany
E-mail: thomas.heine@tu-dresden.de

Dr. Agnieszka Kuc, Prof. Dr. Thomas Heine
Helmholtz-Zentrum Dresden-Rossendorf
Permoserstrasse 15, 04318 Leipzig, Germany

Prof. Dr. Thomas Heine
Department of Chemistry
Yonsei University, Seodaemun-gu, Seoul 120-749, Republic of Korea

**Table S1** – Training data type and composition. G-Phonons refer to phonons being only calculated at the Gamma-point. E stands for total energy, F for forces, and S for stresses. A frame refers to a snapshot of a structure with energy, forces, and stresses, if available.

| Type | Structures | Calculations | Data | Total composition |
| --- | --- | --- | --- | --- |
| Bulk | 1T$^O$, 2H$^O$, 3R$^O$ | Relaxations<br>Phonons<br>Elastic Deformation<br>MD | E+F+S<br>E+F<br>E+F+S<br>E+F | |
| Stacks with 1-3 layers | 1T$^O$, 2H$^O$, 3R$^O$ | Relaxations<br>G-Phonons<br>MD | E+F<br>E+F<br>E+F | 2000 Relaxation frames<br>800 Phonon frames<br>100 Elastic frames<br>9800 MD frames |
| Stacks above 3 layers | 1T$^O$, 2H$^O$, 3R$^O$ | Relaxations<br>G-Phonons | E+F<br>E+F | |
| Ribbons | 1T$^O$ | Relaxations<br>MD | E+F<br>E+F | |
| Surfaces | 1T$^O$ | Relaxations<br>MD | E+F<br>E+F | |

```json
{
  "_comment": "FHI-aims",
  "model": {
    "type_map": ["Pt", "Se"],
    "descriptor": {
      "type": "se_e2_a",
      "sel": [500, 500],
      "rcut_smth": 2.0,
      "rcut": 8.0,
      "neuron": [25, 50, 100],
      "resnet_dt": false,
      "axis_neuron": 12,
      "seed": 1,
    },
    "fitting_net": {
      "type": "ener",
      "neuron": [240, 240, 240],
      "resnet_dt": true,
      "seed": 1,
      "atom_ener": [-518285.968665213, -66543.4947249878],
    },
  },
  "learning_rate": {
    "type": "exp",
    "decay_steps": 5000,
    "start_lr": 0.0001,
    "stop_lr": 3.51e-08,
  },
  "loss": {
    "type": "ener",
    "start_pref_e": 1,
    "limit_pref_e": 1,
    "start_pref_f": 100,
    "limit_pref_f": 1,
    "start_pref_v": 1,
    "limit_pref_v": 1,
  },
}
```

**Figure S1** – Training parameters for DeePMD-kit as defined in the input.json.

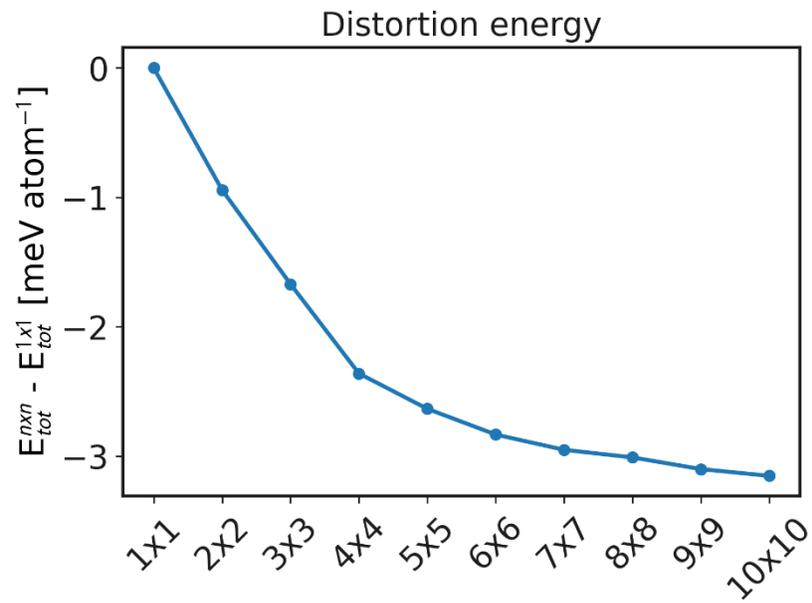

**Figure S2** – Reduction of the total energy of supercells of single-layered PtSe$_2$ after relaxation with the DP model due to small structural distortions, such as wrinkling.

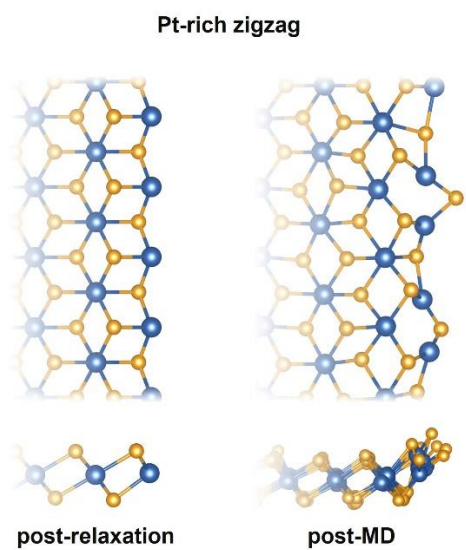

**Figure S3** – Pt-rich zigzag edge after relaxation and after simulated annealing during the DFT Molecular Dynamics (MD) simulation. The Pt-rich edge reconstructs to remove exposed Pt atoms from the edge or surface.

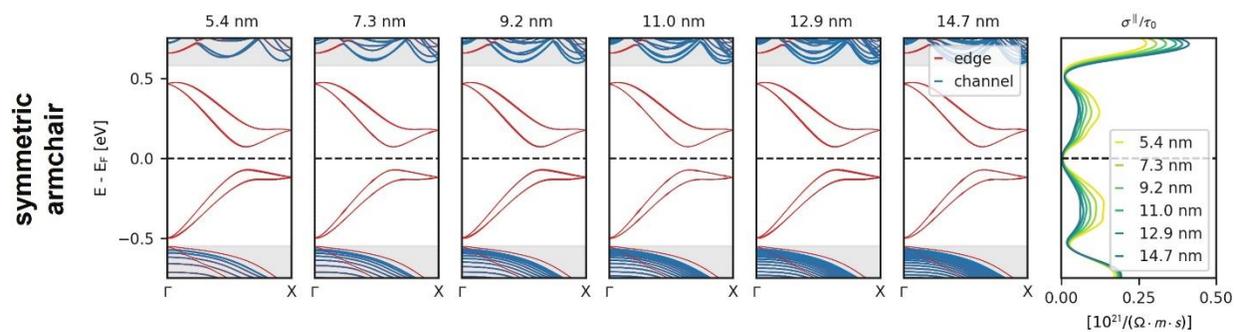

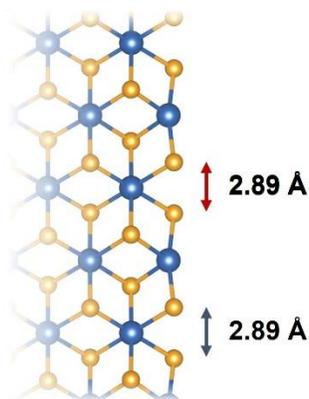 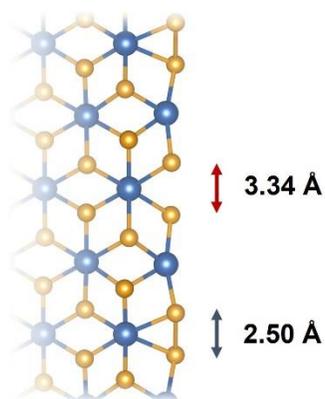

**Figure S4** – Band structure and conductivity of the symmetric armchair edge and illustration how the distorted armchair edge is formed due to dimerization of the Se atoms.

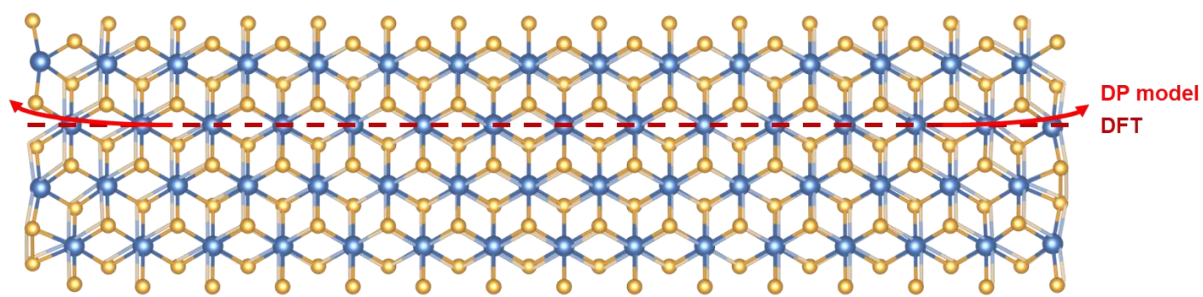

**Figure S5** – Overlay of the distorted armchair ribbon for the smallest width obtained after relaxation via DFT and the DP model.

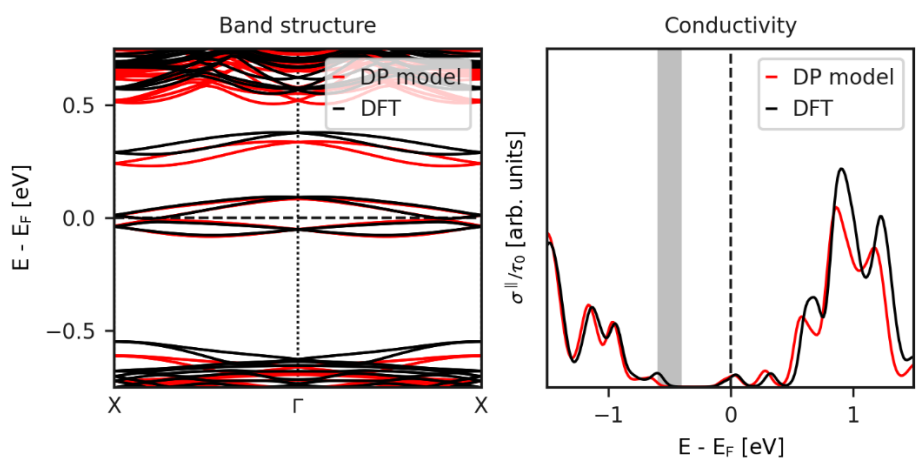

**Figure S6** – Band structure and conductivity of the distorted armchair edge for the smallest ribbon compared between the relaxed structure obtained via DFT and the DP model.

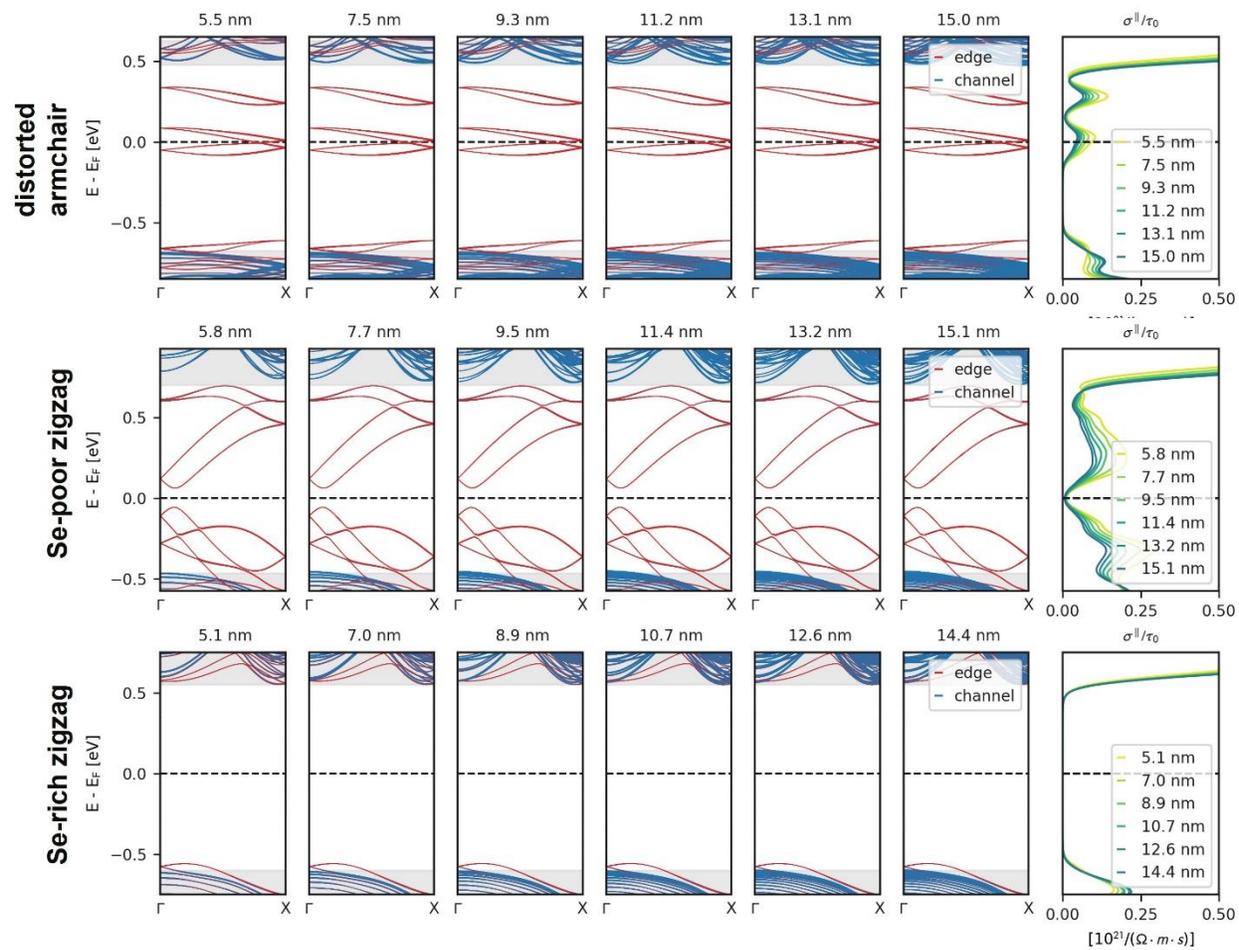

**Figure S7** – Band structures and conductivities for ribbons with different lateral widths and different edge terminations.

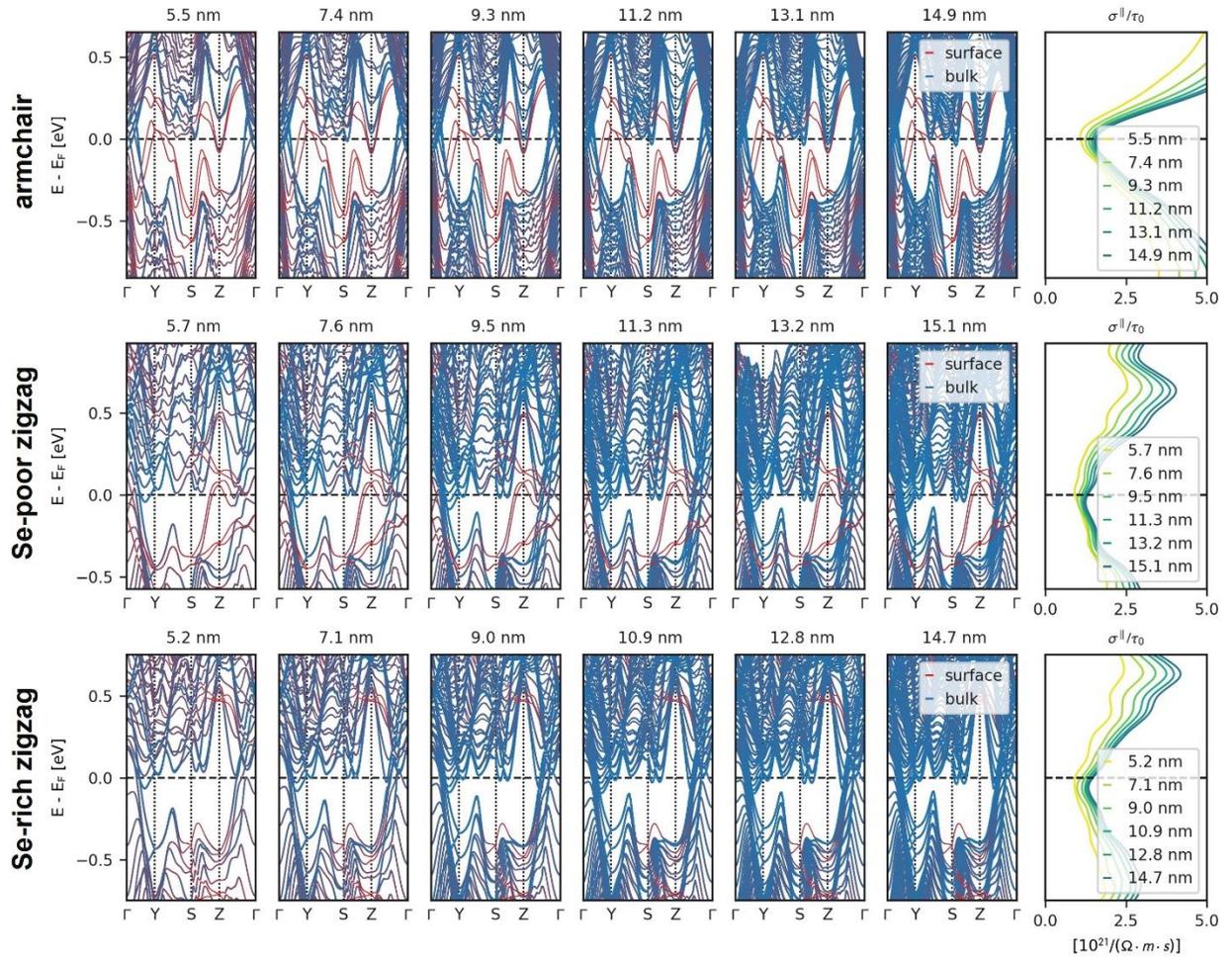

**Figure S8** – Band structures and conductivities for surface stacks with different widths and different edge terminations.

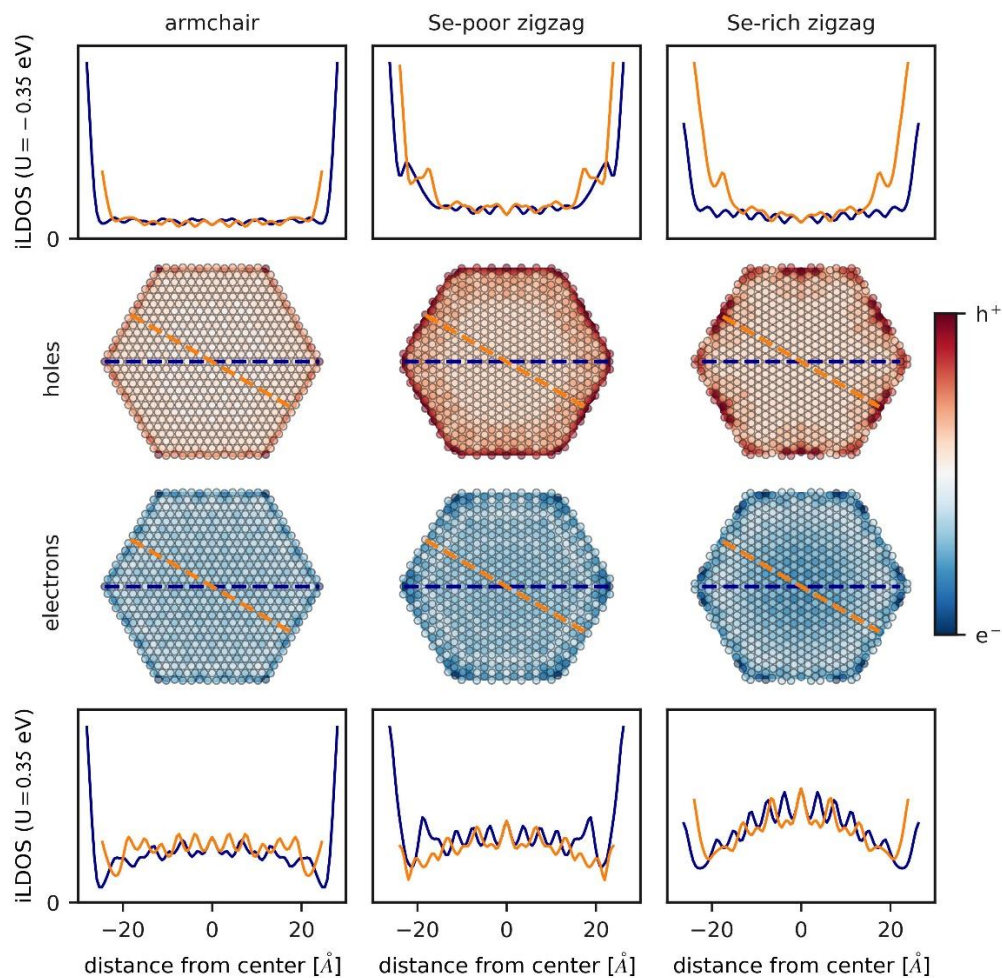

**Figure S9** – Integrated Local Density of States (iLDOS) along line scans and projected onto the basal plane of the nanoplatelets with 5.26 to 5.62 nm lateral width for different edge terminations. The local density of states has been integrated over the holes and electrons in the same energy window as shown in the band structures in main text **Figure 4**.

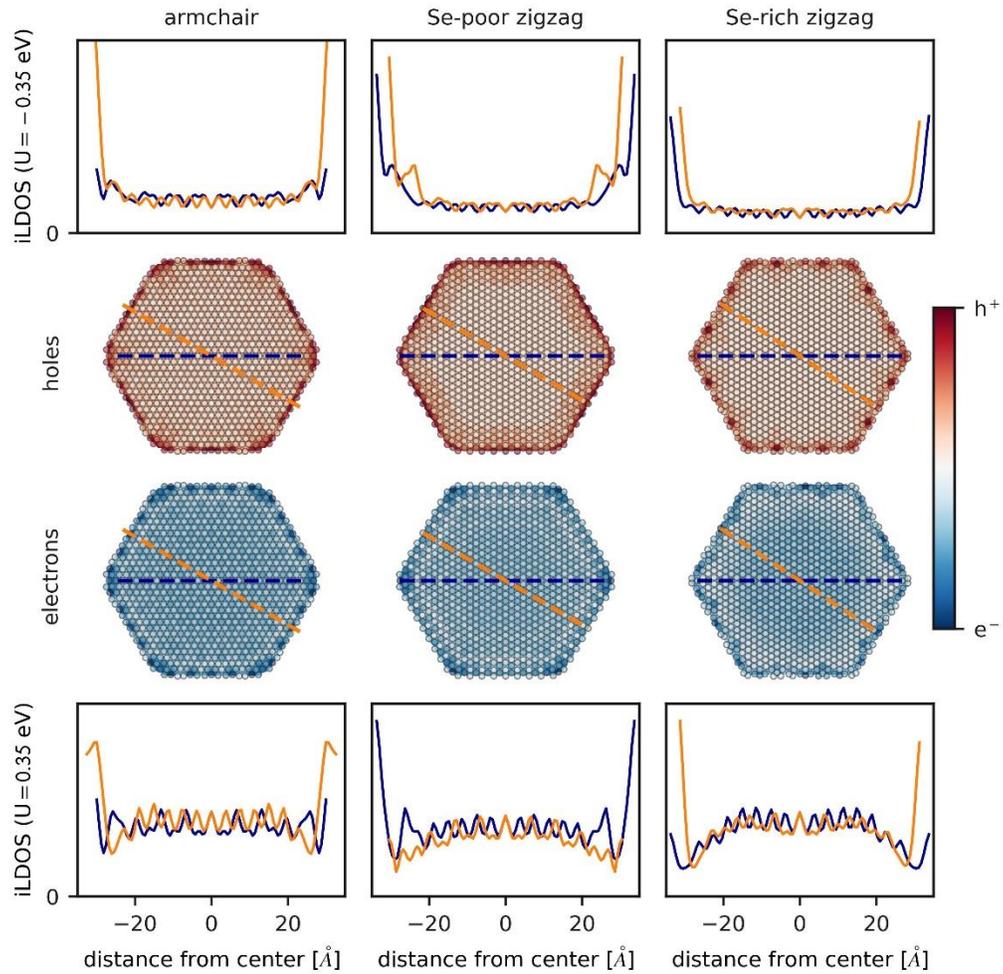

**Figure S10** – Integrated Local Density of States (iLDOS) along line scans and projected onto the basal plane of the nanoplatelets with 6.7 to 7.13 nm lateral width for different edge terminations. The local density of states has been integrated over the holes and electrons in the same energy window as shown in the band structures in main text **Figure 4**.